\documentclass[twocolumn,showpacs,preprintnumbers,amsmath,amssymb,prl,superscriptaddress]{revtex4}

\usepackage{graphicx}
\usepackage{dcolumn}
\usepackage{bm}

\begin{document}

{\bf Reply to the comment\cite{lhcomment2009} on gMechanism of Terahertz Electromagnetic Emission from Intrinsic Josephson Junctionsh\cite{tachiki2009}}\\
\indent We understand that the maximum value of the in-plane superconducting current density caused by the $\pi$-phase kinks is much smaller than the in-plane superconducting critical current density as claimed by the comment. However, this fact does not ensure the stability of the $\pi$-kink state. In our paper we have checked energetically whether the $\pi$-phase kink is stable or not, using the Ginzburg-Landau theory. We found that the reduction of the amplitude of the order parameter around the $\pi$-phase kinks is small but the loss of condensation energy amounts to that comparable to the energy of the kink state which consists of super-current energy and electric and magnetic ones. This result comes from the fact that the superconducting condensation energy in the superconducting CuO$_{2}$ layers in Bi$_{2}$Sr$_{2}$CaCu$_{2}$O$_{8}$ (BSCCO) is very large. On the basis of this estimation we concluded that the kink state is unfavorable compared with the state without kink.  To criticize our conclusion a quantitative argument is required.\\
\indent  We made a frequency analysis of the time evolution of the oscillating electric field shown in Fig. 5(a) of Ref. \cite{tachiki2009} and found that there is a sub-harmonic component with one half of the Josephson frequency given by $(2e/h)\nu$, $\nu$ being the voltage between the superconducting CuO$_{2}$ layers in BSCCO.  This sub-harmonic in the state without kink seems to arise from a parametric excitation caused by the non-linearity of this system \cite{fulton1973}. The calculated phase difference exhibits just a standing wave oscillation and no sign of propagating soliton mode as stated in the comment, which appears in a single long Josephson junction.  We admit that the existence of a sub-harmonic component is not consistent with the result of the frequency analysis of experiments \cite{oz2007} and, therefore, our theory should be improved by taking account of some mechanism which suppresses the sub-harmonic.  However, this fact does not lead to the conclusion that the THz emission observed in the experiment comes from the $\pi$-kink state, though the frequency of the fundamental excitation mode in the $\pi$-kink state certainly fulfills the Josephson relation in consistence with the experiments, because the $\pi$-kink state has another serious difficulty. As discussed below, the strong emission cannot be expected in the $\pi$-kink state. To obtain the $\pi$-kink solution one has to impose the boundary condition that the oscillating magnetic field is extremely weak at the sample side surfaces \cite{lh2008, hl2008, lh2009}, which indicates that the electromagnetic waves are not emitted from the side surfaces, because the Poynting vector has nearly zero value at the boundary. Then, to evaluate the intensity of the far-field radiation from the $\pi$-kink state a formula for the patch antenna in the antenna theory \cite{leone2003} is utilized in the papers given in Refs. \cite{lh2009} and \cite{leone2003}. It is noted that the formula describes the power emitted from normal carriers oscillating parallel to the in-plane direction inside the electrodes that cover the top and bottom junctions. The use of the formula may not be justified for the systems such as BSCCO mesas in which a strong AC tunneling current can flow between the electrodes. We have numerically checked, using the $xz$ model presented in the recent paper by Koyama {\it et~al.} \cite{koyama2009}, that the flow of normal carriers along the in-plane direction is negligibly small when the electrode is a metal with large electric conductivity. We have also observed that the surface impedance at the side surfaces depends on the voltage appearing in the junctions in the $xz$ model.  The impedance at the voltage at which strong emission takes place is decreased several times from the value in the off-resonant state. These results indicate that the assumptions that the AC magnetic field is negligibly small at the side surfaces of the junctions and the emission arises from the in-plane current are not justified.\\
\indent As mentioned above, all the approaches up to now have not been successful in explaining the systematics of the experimental results. In order to obtain the complete answer to the mechanism of THz emission in intrinsic Josephson junctions, studies using a more realistic model containing electrodes and the space outside the junctions as those presented in Ref. \cite{koyama2009} will be required.\\
\indent  This work has been supported by the JST (Japan Science and Technology Agency) CREST project.\\
\\\\
Masashi Tachiki$^{1}$, Tomio Koyama$^{2}$ and Shouta Fukuya$^{1,3}$\\\\
$^{1}$Institute of Material Science, Universioty of Tsukuba, Tsukuba 305-8577, Japan\\
$^{2}$Institute for Materials Research, Tohoku University, Sendai 980-8577, Japan\\
$^{3}$Institute for Solid State Physics, University of Tokyo, Kashiwa 277-8581, Japan


\end{document}